# ION CHANNELS, NATURAL NANOVALVES

Bob Eisenberg
Rush University, Chicago, IL USA
beisenbe@rush.edu

*As prepared for*

**Topical Editor**

# Werner Kunz

Institute of Physical and Theoretical Chemistry
University of Regensburg,
D-93040 Regensburg

*and the*

# Encyclopedia of Applied Electrochemistry

*Springer*

### Editors-in-Chief

**Robert F. Savinell**
Case Western Reserve University, Cleveland
Former North American Editor of the Journal of Applied Electrochemistry

**Ken-ichiro Ota**
Yokohama National University
President of the Electrochemical Society of Japan

**Gerhard Kreysa**
Technical University of Clausthal
Former Chief Executive of DECHEMA

# Main Text

Ion channels are proteins with holes down their middle that control the flow of ions and electric current across otherwise impermeable biological membranes (Hodgkin 1992; Sakmann and Neher 1995; Ashcroft 1999; MacKinnon 2004). The flow of $Na^+$, $K^+$, $Ca^{2+}$, and $Cl^-$ ions have been central issues in biology for more than a century. The flow of current is responsible for the signals of the nervous system that propagate over long distances (meters). The concentration of $Ca^{2+}$ is a 'universal' signal that controls many different systems inside cells. The concentration of $Ca^{2+}$ and other messenger ions has a role in life rather like the role of the voltage in different wires of a computer. Ion channels also help much larger solutes (e.g., organic acid and bases; perhaps polypeptides) to cross membranes but much less is known about these systems.

Ion channels can select and control the movement of different types of ions because the holes in channel proteins are a few times larger than the (crystal radii of the) ions themselves. Biology uses ion channels as selective valves to control flow and thus concentration of crucial chemical signals. For example, the concentration of $Ca^{2+}$ ions determines whether muscles contract or not. Ion channels have a role in biology similar to the role of transistors in computers and technology(Eisenberg 2012). Ion channels control concentrations important to life the way transistors control voltages important to computers.

Ion channels are not always open. Single channel molecules switch suddenly (in less than one microsecond) between definite open and closed states. Single channel molecules open and close stochastically (in a process called spontaneous 'gating') according to well defined probability distributions (Sakmann and Neher 1995) that are usually well described by Markov models (with a few rate constants that depend dramatically on conditions (Eisenberg 2011)).

The biological properties of channels are determined by the properties of the ensemble of channels. The gating properties of the ensemble of channels is not noticeably stochastic. The properties of the ensemble of voltage dependent channels in nerve fibres are well described by partial differential equations (Hodgkin 1958) invented to allow computation of the propagating voltage signal of nerve fibers (Hodgkin 1992). Different equations are needed for different ions (e.g., $Na^+$ and $K^+$) because the ions flow through different proteins that have different structures and properties. The currents through these different channel proteins combine according to the same equations used to describe transmission lines in engineering, the equations first used by Kelvin to describe the spread of telegraph signals in an insulated cable under the Atlantic ocean. The solutions of the cable equations of Hodgkin and Huxley are the signals of the nervous system. The signals arise from the coupling of macroscopic boundary conditions and conservation laws (of the transmission line equations) to the gating properties of ensembles of channels (described by the Hodgkin Huxley equations) and the selectivity properties of the hole in single channel proteins (see below). The biological function of nerve is understood and computed by a multi-scale analysis, with each scale linked in an explicit and calibrated way to its neighbouring scales. It is likely that such multiscale analysis will be needed for many biological systems.

Different types of channels gate in response to different stimuli because different channel proteins have different structural modules to sense the stimuli. Some channels have modules that respond to voltage. Others have modules that respond to temperature. Still others have modules that respond to specific chemicals, e.g., capsaicin, the ingredient that makes chili peppers taste hot. New channels with new types of sensitivity are constantly being discovered. The diversity is remarkable (Abelson, Simon, Ambudkar and Gottesman 1998). A substantial fraction of all proteins in a human are ion channels.

Ion channels are so important biologically and medically that they are studied in thousands of laboratories every day by molecular biologists and physicians interested in understanding and controlling disease. Many of the most important drugs in clinical medicine work by controlling the gating of channels. A common strategy of the pharmaceutical industry is to identify agents that control channels and then see what those agents do clinically.

Ion channels are proteins and so can be identified and classified by the powerful methods of molecular and structural biology. These methods are used throughout the worlds of biotechnology and pharmaceutical science, as well as widely in medical and clinical science. These methods readily provide information with atomic resolution not available for most non-crystalline physical systems. The location of individual atoms are known in thousands of different proteins. Remarkably, the chemical nature of small groups of atoms (the side chains of amino acids that make up the protein) can be easily changed by the technique of site-directed mutagenesis. Such mutation experiments often discover small groups of side chains that control the biological function of proteins and channels. The genome (and evolution) evidently control function in this way.

The selectivity of ion channels is often determined by only a handful of amino acids. Thus the amino acids Glu-Glu-Glu-Glu (EEEE) determine the selectivity of the calcium channel of the heart (Sather and McCleskey 2003); the amino acids Asp-Glu-Lys-Ala (DEKA) determine the selectivity of the sodium channel of nerve and muscle cells (Payandeh, Scheuer, Zheng and Catterall 2011). The structure of ion channels is notoriously difficult to determine because channel proteins do not easily crystallize. They are normally found in lipid membranes and so methods suitable for crystallizing soluble proteins are not too helpful. Several structures have been determined, most notably the KcsA potassium channel (MacKinnon 2004).

Ion channels are unusual proteins. The current through a single channel is the same whether the channel is open a very short time (microseconds) or a very long time (seconds). We conclude that an ion channel does not change structure significantly, once it is open.

Ion channels are unusual proteins because an important part of their function occurs without conformational changes. Current flow through ion channels is driven by electrodiffusion through a hole of one conformation, once the channel opens. Thus, an open ion channel is a physical object that can be analysed by the techniques of classical physical chemistry. We do not need to understand vaguely defined conformation changes or mysterious allosteric effects. We do not need to understand the opening process to understand the open channel.

Ion channels allow atomic scale structures to control macroscopic function so it is natural to seek understanding with models that include all atomic detail using the methods of molecular dynamics to compute atomic motion. The problem with this approach is that atoms move a great deal, at the speed of sound, as a first approximation (Berry, Rice and Ross 1963), and so computations must resolve $10^{-16}$ sec. Little biology happens faster than $10^{-4}$ seconds. Atomic scale computations must extend over 12 orders of magnitude in time if they are to compute biological function.

Current flow through channels depends on bulk concentrations of ions ranging from $10^{-7}$M (or even much less) to 0.5 M. Concentrations of ions in and near ion channels (and in and near active sites of enzymes) are much higher, even larger than 10M because the systems are so small and have large densities of permanent charge from acid and base side chains of proteins. Computations of these concentrations must include nonideal properties of highly concentrated interacting solutions and side chains because these are known to be of great importance in ionic mixtures or solutions greater than 50 mM concentration. Simulations in full atomic detail must also include staggering numbers of water molecules to deal with the trace concentrations of $10^{-11}$ to $10^{-7}$ M of signalling ions, e.g., $Ca^{2+}$.

Molecular dynamics of biological function must be done in nonequilibrium conditions where flows occur, because almost all biology occurs in these conditions. Simulations have difficulty dealing with the action potentials of nerve and muscle fibers. Molecular dynamics cannot compute the billions of trajectories of ions that cross membranes to make action potentials, lasting milliseconds to nearly a second, flowing centimeters to meters down nerve axons. Simulations at present cannot deal with flows that are controlled by channels on the atomic scale but couple to boundary conditions on the macroscopic scale, millimeters away from the channel. Many biological systems use atomic scale structures this way to control macroscopic function.

Simulations must deal with all these issues of scale at once, because biology uses them all at once. For these reasons, molecular dynamics cannot deal directly with biological function, as of now. A multiscale approach with explicit models and calibrated links between scales seems unavoidable, as in the analysis of the propagating voltage signal of nerve fibres, and in engineering technology (Eisenberg 2011), in general.

Multiscale reduced models of ion channels are feasible. Reduced models of some open ionic channels have proven surprisingly successful (Eisenberg 2011). In several cases, it has been possible to understand, and predict experimental results before they were done. Channels have been built that behave as expected (Vrouenraets, Wierenga, Meijberg and Miedema 2006). These reduced models include surprisingly little atomic detail; for example, they treat water as a continuum dielectric and side chains as spheres. It is not clear why a sensitive biological function like selectivity can be explained with such little regard to the atomic details of hydration and solvation, but the evidence is clear. In several important cases, the explanation is successful.

Specifically, the selectivity of the ryanodine receptor of cardiac and skeletal muscle can be understood with a model with less than a dozen parameters that never change value (Gillespie 2008). Detailed properties of the current through the channel were successfully predicted (in quantitative detail, with errors of a few per cent) with this model, often before the experiments were performed. Predictions were successful after drastic mutations, and in many (>100) solutions, of widely varying composition.

The calcium channel of cardiac muscle has a complex pattern of binding of ions that can be understood (over four orders of magnitude of concentration in many types of solutions, containing $Na^+$, $K^+$, $Rb^+$, $Cs^+$, $Ca^{2+}$, $Ba^{2+}$, $Mg^{2+}$, and so on) with a model containing two or three parameters (Boda, Valisko, Henderson, Eisenberg, Gillespie and Nonner 2009). Specific mutations change this model of a calcium channel into a sodium channel (Boda, Nonner, Valisko, Henderson, Eisenberg and Gillespie 2007) just as the mutations change the selectivity in experiments. A reduced model can explain these data. The model has just three parameters that never change value, namely the diameter of the channel, the dielectric coefficient of the solution, and the dielectric coefficient of the protein. Ion diameters in the model never change value. Reduced models of this type have not yet accounted for the selectivity of potassium channels and it is not clear if they can.

## Future Directions

Biologists and physicians will continue to discover and describe the thousands of types of channel proteins that make life possible. Physical scientists will seek models that include enough atomic detail to explain the role of structure, while mathematicians and computer scientists struggle to deal with the motions of those atoms and other problems of scale. Physical chemists will seek simple principles that control the selectivity of ion channels and will try to apply those principles to electrochemical systems of technological interest. The mystery of gating will eventually be resolved. Perhaps, the steady time independent current of a single open channel will emerge from the analysis of the stability of a coupled ion, channel, bath system, as some type of nonlinear 'eigenstate'. Biologists will study information from the multiplicity of channels, seeking the pattern(s) that evolution has used to create function, from structure, and physics.

Physical scientists will approach main questions: How can multiscale calibrated models of ion channels be built, computed, and tested? How do specific structures control the opening and closing of channels? How do proteins sense voltage? What is opening and closing when channels open?, Why do reduced models of selectivity work for some channels, and not others? How can we increase the flow of current through channels so they perform better? How can we mimic the properties of biological channels in technologically useful systems?

Ion channels are unique objects: they have enormous biological and clinical importance. They use simple physical forces to perform those functions. Ion channels are likely to be one of the first important biological systems that can be understood in full detail, in the tradition of physical science. Understanding ions in channels may be as important to the future of mankind as understanding holes and 'electrons' in semiconductors has been to our recent past.

## Cross-References

Specific Ion Effects: Evidences; Cell Membranes; Bio-electrochemical Energy Transformation.